 \documentclass[twocolumn,aps,prb,superscriptaddress,10pt]{revtex4-1}	
\usepackage{amsmath,amsfonts,amsthm}			
\usepackage{adjustbox,lipsum}
\usepackage{subcaption}
\usepackage[format=plain,indention=-0.3cm,labelfont={bf,normalsize},font={small},justification=raggedright]{caption}
\usepackage{float}
\usepackage[caption=false]{subfig}
\usepackage{url}    
\usepackage{epsfig}
\usepackage{color}

\begin{document}
\title{Nonempirical Range-separated Hybrid Functionals for Solids and Molecules}
\author{Jonathan H Skone}
\affiliation{Institute for Molecular Engineering, University of Chicago, 5801 South Ellis Avenue, Chicago, IL 60637}
\affiliation{Materials Science Division, Argonne National Laboratory, Argonne, IL 60439}
\author{Marco Govoni}
\affiliation{Institute for Molecular Engineering, University of Chicago, 5801 South Ellis Avenue, Chicago, IL 60637}
\affiliation{Materials Science Division, Argonne National Laboratory, Argonne, IL 60439}
\author{Giulia Galli}
\email[Author to whom electronic correspondence should be addressed. Electronic address: ]{gagalli@uchicago.edu}
\affiliation{Institute for Molecular Engineering, University of Chicago, 5801 South Ellis Avenue, Chicago, IL 60637}
\affiliation{Materials Science Division, Argonne National Laboratory, Argonne, IL 60439}
\date{\today}
\begin{abstract}
Dielectric-dependent hybrid (DDH) functionals were recently shown to yield accurate energy gaps and dielectric constants for a wide variety of solids, at a computational cost considerably less than that of GW calculations. The fraction of exact exchange included in the definition of DDH functionals depends (self-consistently) on the dielectric constant of the material. Here we introduce a range-separated (RS) version of DDH functionals where short and long-range components are matched using system dependent, non-empirical parameters. We show that RS DDHs yield accurate electronic properties of inorganic and organic solids, including energy gaps and absolute ionization potentials. Furthermore we show that these functionals may be generalized to finite systems.
\end{abstract} 

%
\maketitle 
%

\section{ \textbf {Introduction}}
\label{intro}

To hasten the discovery of new materials with optimal properties for applications such as optical and electronic devices, catalysis, quantum information, and photovoltaics, the reliance on theory to support and direct experimental efforts is essential. The determination of the ground and excited state electronic properties relevant to such applications, require a high level of accuracy, ideally at a modest computational cost. Density functional theory (DFT)\cite{Gross:DFTbook} has long been one of the main methodologies of choice as it provides a reasonable compromise between accuracy and computational efficiency. Some of the most accurate functionals include an admixture of local and nonlocal exchange and are referred to as hybrid functionals.\cite{Barone:1994hq} The latter have been widely used for molecules and less so to model extended systems.  
The slower adoption of hybrid functionals for condensed phases is a consequence of using plane-wave basis sets in most condensed matter calculations, and of the lack of nonempirical parameters for constructing accurate hybrid functionals.\footnote{The basis representation common to the vast majority of quantum chemistry codes is instead localized basis sets (typically atom centered Gaussian functions)} 
Indeed the treatment of the non-local exchange operator within periodic boundary conditions by using a plane-wave basis set is computationally demanding. In the last decade, however, due in part to several methodological advances,\cite{Wu:2009ub,Gygi:2009jn, Guidon:2010jc, Gygi:2013ic} hybrid functionals have been increasingly used to investigate a variety of periodic systems within a plane-wave pseudopotential framework. 

Addressing the need for improved accuracy of DFT for condensed systems, a new class of functionals was recently proposed, with parameters defined using the dielectric constant of the system.\cite{Shimazaki:2008tr,Shimazaki:2009hx,Marques:2011fi,Koller:2013ht,Skone:2014hu,RefaelyAbramson:2013bv} This class of functionals, referred to as dielectric-dependent hybrid (DDH) functionals, can yield accurate electronic structures of solids, at considerably less cost with respect, e.g. to GW based methods.
 Among DDHs, is one in which the dielectric constant is determined self-consistently,\cite{Shimazaki:2008tr,Shimazaki:2009hx,Skone:2014hu} which is referred to as a sc-hybrid.\cite{Skone:2014hu}
This sc-hybrid functional has found recent use in the study of pristine oxides,\cite{Pacchioni:2014dk,Skone:2014hu, Gerosa:2015gt} defects in oxides,\cite{Gerosa:2015eb, Gerosa:2015uo, Gerosa:2016WO3} nitrides,\cite{Morbec:2014je, Morbec:2016dz} aqueous solutions,\cite{Gaiduk:2016} and spin-defects in wide band gap semiconductors.\cite{Seo:2016}
The adaptation of non self-consistent DDHs to a time-dependent framework was recently pursued by Yang \textit{et al.},\cite{Yang:2015dj} Ferrari \textit{et al.},\cite{Ferrari:2015ik} and Refaely-Abramson \textit{et al.},\cite{Refaely-Abramson:2015} with the goal of obtaining optical spectra of solids.
Another recent study by Shimazaki \textit{et al.}\cite{Shimazaki:2015eu} explored the possibility of defining a {\it local} DDH functional where the fraction of exact-exchange is site dependent. Though their initial prescription using atom-centered basis functions is perhaps oversimplified, the proposal of Ref~\onlinecite{Shimazaki:2015eu} is an interesting step towards describing {\it heterogenous systems} with DDH functionals.

In this paper we introduce a range-separated (RS) version of DDH functionals where short and long-range components are matched using system dependent, nonempirical parameters.
We assess the accuracy of RS-DDH functionals for the electronic properties of inorganic materials and molecular crystals, and we present calculations for energy gaps and absolute ionization potentials. We show that range-separated DDH functionals are superior to full-range DDH functionals for the energy gaps of inorganic solids, while both full and range separated hybrids yield equally accurate results for the gaps and ionization potentials of molecular crystals. In addition, we show that RS-DDH can be generalized to finite systems.

The rest of the paper is organized as follows. Section \ref{methods} describes the range-separated DDH functional along with the computational details used in this work. Section \ref{results} presents results obtained using nonempirical parameters for the range-separated and full-range DDH functionals for a set of inorganic solids, molecular crystals, and finite systems. Section \ref{summary} summarizes our results, and provides our conclusions. 

\section{ \textbf {Method}}
\label{methods}
\subsection{Range-separated dielectric-dependent hybrid functionals}

We use a Generalized Kohn Sham (GKS) framework where we determine an effective screening of the Coulomb potential by computing the dielectric response of the system. \\

The GKS nonlocal potential $ v_{_{\mathrm{GKS}}}(\mathbf{r,r'}) $ entering the Kohn-Sham (KS) Hamiltonian is given by:
\begin{equation}
 v_{_{\mathrm{GKS}}}(\mathbf{r,r'}) =  v_H(\mathbf{r}) + v_{x}(\mathbf{r,r'}) + v_{c}(\mathbf{r})  + v_{ext}(\mathbf{r})
\label{equation:GKS}
\end{equation}
where $v_{H}(\mathbf{r})$ is the Hartree, $v_{x}(\mathbf{r,r'})$ is the nonlocal exchange potential, $v_{c}(\mathbf{r})$ is the correlation potential, and $v_{ext}(\mathbf{r})$ is the attractive Coulomb potential to the nuclei. \\

The nonlocal exchange potential $v_{x}(\mathbf{r,r'})$ is partitioned into long-range (lr) and short-range (sr) components where $\alpha$ and $\beta$, define the fraction of exact exchange admixed to semilocal exchange in the lr and sr components, respectively: 
\begin{eqnarray}
v_{x}(\mathbf{r,r'}) &=&  \alpha v_x^{\mathrm{lr-ex}}(\mathbf{r,r';\mu}) + \beta v_x^{\mathrm{sr-ex}}(\mathbf{r,r';\mu}) \label{equation:GKSxc} \\
            \nonumber                 && + (1- \alpha )v_x^{\mathrm{lr}}(\mathbf{r; \mu}) + (1-\beta)v_x^{\mathrm{sr}}(\mathbf{r; \mu})  \, .
\label{equation:RSHeqn}
\end{eqnarray}
The screening parameter $\mu$ defines how the lr and sr components are bridged. We use the error function to define the range separation of the Coulomb interaction, namely: 

\begin{equation}
v_x^{\mathrm{lr-ex}}(\mathbf{r,r'; \mu}) = -\rho(\mathbf{r},\mathbf{r'}) \frac{\mathrm{erf}(\mu \mathbf{|r-r'|})}{|\mathbf{r-r'}|}
\label{equation:exact-exchange-lr}
\end{equation}
and
\begin{equation}
v_x^{\mathrm{sr-ex}}(\mathbf{r,r'; \mu}) = -\rho(\mathbf{r},\mathbf{r'}) \frac{\mathrm{erfc}(\mu \mathbf{|r-r'|})}{|\mathbf{r-r'}|} \, , 
\label{equation:exact-exchange-sr}
\end{equation}

where $\rho(\mathbf{r},\mathbf{r'})$ is the density matrix. The semilocal exchange potentials, $v_x^{\mathrm{sr}}(\mathbf{r; \mu})$ and $v_x^{\mathrm{lr}}(\mathbf{r; \mu})$, depend only on the density $\rho(\mathbf{r})$ and its gradient. 
Here we adopt the PBE\cite{Perdew:1996ug} approximation of the semilocal exchange $v_{x}(\mathbf{r})$ and correlation $v_{c}(\mathbf{r})$. 
Many of the commonly used exchange functional forms may be recovered from Eq.~\ref{equation:GKSxc}. For example, when $\alpha = \beta = \mu = 0$, one obtains the PBE semilocal functional. If $\alpha = 1$, $\beta = 0$, and $\mu \to \infty$ one obtains the KS equations with the exact-exchange potential (EXXc\cite{MoriSanchez:2006gl}). If instead $\alpha = 0.25$, $\beta = 0$, and $\mu \to \infty$ the PBE0 hybrid functional is recovered. Short-range hybrid functionals ($\alpha=0$) may also be easily obtained e.g. HSE06\cite{Heyd:2006dc} where $\beta=0.25$ and $\mu=0.11$ bohr$^{-1}$ or sX-LDA\cite{Bylander:1990we} where $\beta=1$, $\alpha=0$, and the Thomas-Fermi screening function is used instead of the error function\footnote{With sX-LDA the local exchange and correlation potential used is LDA.}. Examples of long-range hybrid functionals ($\alpha \ne 0$) include the empirical CAM-B3LYP functional,\cite{Yanai:2004we} where $\alpha=0.46, \beta=0.19, \mu=0.33$ bohr$^{-1}$, as well as LC-$\mu$PBE,\cite{Weintraub:2009ub} where $\alpha=1$, $\beta=0$, and $\mu=0.4$ bohr$^{-1}$. The screened-exchange methodology of Robinson \textit{et al.}\cite{Robinson:1962tp} and subsequent works,\cite{Bylander:1990we, Clark:2010ed, Guo:2014fs} present a similar approach to the description of electronic screening, but the function used to partition the Coulomb operator and the asymptotic long-range limit of the Coulomb potential differ from those used here. \\

The goal of the present work is to generalize the self-consistent dielectric-dependent hybrid functional (sc-hybrid)\cite{Skone:2014hu} to a range-separated form. 
To do so we first compare results obtained with the PBE0 global hybrid ($\alpha=0.25$) and the sc-hybrid, where the global fraction of exchange is inversely proportional to the self-consistently determined electronic dielectric constant ($\epsilon_\infty$), namely
\begin{equation}
\alpha = \frac{1}{\epsilon_\infty}.
\label{equation:alpha_epsilon}
\end{equation}

\begin{figure}[]
\centering
\scalebox{0.3}{\includegraphics[trim= 0mm 0mm 0mm 0mm,clip, angle=0]{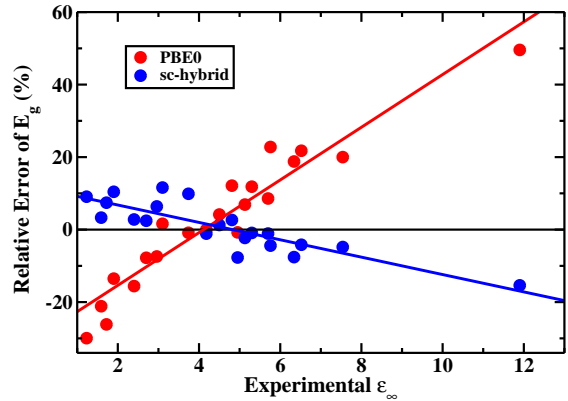}}
\caption{(Color online) Plot of the signed relative error in the electronic gap ($E_g$) for a set of semiconductors and insulators (see Table~\ref{table:Eg-comp-tab}) computed at the PBE0 (red circles and line) and sc-hybrid (blue circles and line) levels of theory as a function of the experimental electronic dielectric constant $\epsilon_{\infty}$. The CoO material was removed from this analysis since the sc-hybrid functional appears to be ill-suited for systems with localized d-electrons at the valence band edge.}
\label{fig:rel-error-gap}
\end{figure}

As discussed in Ref.~\onlinecite{Skone:2014hu}, Eq.~\eqref{equation:alpha_epsilon} can be obtained from the static COHSEX self-energy,\cite{Hedin:1965hi} by approximating the screened Coulomb interaction $W$ with an effective screened interaction, where the inverse microscopic dielectric function is replaced by the inverse macroscopic dielectric constant: 
\begin{equation}
W(\mathbf{r},\mathbf{r'}) =
\int d{\mathbf{r''}}  \epsilon^{-1} (\mathbf{r},\mathbf{r''}) v(\mathbf{r''},\mathbf{r'})
\approx 
 \frac{\alpha}{|\mathbf{r}-\mathbf{r'}|} \,.
\label{eq:wstat}
\end{equation}
 
We previously found that for a set of diverse semiconductors and insulators the mean absolute relative error (MARE) in the computed energy gaps was $\sim 17 \%$ and $\sim 7.0 \%$ when using PBE0 and the sc-hybrid, respectively. 
Fig.~\ref{fig:rel-error-gap} shows the signed relative error of the predicted electronic gaps with respect to experiment, as a function of the experimental dielectric constant. For $\epsilon_\infty > 4$ the PBE0 (sc-hybrid) over- (under-) estimates the experimental energy gap, and vice versa for $\epsilon_\infty < 4$. Hence for systems with $\epsilon^{-1}_\infty = \alpha$ $<$ ($>$) 0.25, increasing (lowering) $\alpha$ from 0.25 may improve the agreement with expereimental photoemission gaps.  

Based on this observation, we defined a new class of DDH functionals having the general form of Eq.~\eqref{equation:GKSxc}, where we set the long range fraction $\alpha=\epsilon_\infty^{-1}$ and the short range fraction $\beta=0.25$ (as in PBE0). Within this framework, the approximate $W(\mathbf{r},\mathbf{r'})$ becomes:
\begin{equation}
W(\mathbf{r},\mathbf{r'}) 
\approx 
\frac{\epsilon^{-1}_{\infty}}{|\mathbf{r}-\mathbf{r'}|} +  ( \beta - \epsilon^{-1}_{\infty} ) \frac{\mathrm{erfc}(\mu \mathbf{|r-r'|})}{|\mathbf{r-r'}|} ,
\label{eq:wstatRSH}
\end{equation}
where the first term on the right hand side of Eq.~\eqref{eq:wstatRSH} is the same as the sc-hybrid functional of Ref.~\onlinecite{Skone:2014hu} and the second term is a short-range correction to the Coulomb potential. 
The numerator of Eq.~\eqref{eq:wstatRSH} is plotted as a function of $\mu$ in Fig. S1 of the Supplemental Material.
Note that the expression of Eq.~\eqref{eq:wstatRSH} is general, and $\beta$ is in principle a parameter to be determined; here we chose $\beta = 0.25$ based on the results of Fig.~\ref{fig:rel-error-gap}. 
The plot indicates that a value of $\mu$ in the range ($0, +\infty)$ exists which may improve the description of the electronic gaps with respect to both PBE0 and sc-hybrid functionals. 
We note that the long-range limit of the  RS DDH functional defined here is physically correct; however the short-range limit of the exchange is not correctly unscreened, but rather  attenuated by the PBE0 fraction of exchange (0.25). This attenuation factor of the exchange amounts, in practice, to introducing an approximate form of short-range correlation in the functional.

We explored three different non-empirical choices of $\mu$ which do not require any optimization procedure, nor calculations for charged systems, unlike e.g. other definitions of range-separated hybrid functionals,\cite{Baer:2010gf,RefaelyAbramson:2012uw,Korzdorfer:2012jx} or Koopman complaint functionals.\cite{Dabo:2010jc,Poilvert:2015fv,Nguyen:2015if} 

\begin{figure}[]
\centering
\scalebox{0.35}{\includegraphics[trim= 0mm 0mm 40mm 0mm,clip, angle=0]{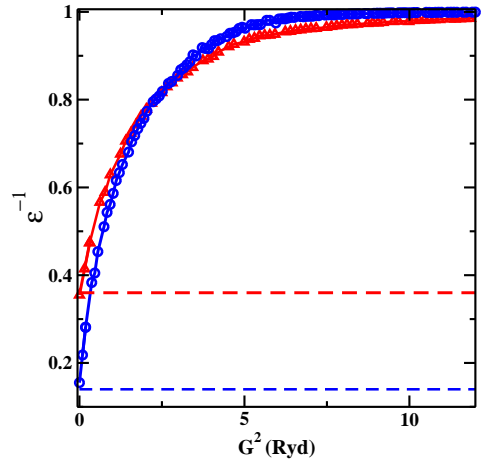}}
\caption{(Color online) Plot of the inverse diagonal dielectric function ($\epsilon^{-1}$) in Fourier space as a function of G$^2$ for a prototypical insulator MgO (red triangles and line) and a prototypical small gap semiconductor AlP (blue circles and line). The dashed lines correspond to $\epsilon_{\infty}^{-1}$ screening that is seen over all inter-electronic distances for the global sc-hybrid.} 
\label{fig:inv-die-funct-mgo}
\end{figure}

(I.) Assuming that a homogeneous spatial distribution of the valence electrons is a good approximation of the valence charge density of the system, we defined a screening length related to the average volume occupied by a valence electron:
\begin{equation}
\mu_{WS} = \frac{1}{r_s} = \left( \frac{4\pi n_v}{3} \right)^{\frac{1}{3}}
\label{eq:WS}
\end{equation}
where $n_v$ is the valence electron density, and $r_s$ is the Wigner-Seitz radius. 

(II.) Alternatively, we used the Thomas-Fermi screening parameter ($\mu_{\text{TF}}$):

\begin{equation}
\mu_{\text{TF}} = \frac{1}{2}k_{TF} = \left(\frac{3 n_v}{\pi}\right)^{\frac{1}{6}}
\label{eq:TFsimplify}
\end{equation} 
where $k_{TF} = 2 \left(\frac{3 n_v}{\pi}\right)^{\frac{1}{6}}$ is the Thomas-Fermi screening length. 
Note that in the definition of both $\mu_{WS}$ and $\mu_{TF}$, atomic units are used,\footnote{$a_0 = (\frac{\hbar^2}{me^2})$} and $\mu_{\text{TF}} =0.781593\sqrt{\mu_{\text{WS}}}$.

(III.) Finally we defined a range separation parameter $\mu$ obtained from the long-range decay of the diagonal elements of the dielectric matrix, as computed from first principles using the linear response techniques proposed in Ref.~\onlinecite{Wilson:2008jp} and implemented in the \texttt{WEST} code.\cite{Govoni:2015gc} Fig.~\ref{fig:inv-die-funct-mgo} shows the dielectric function  ($\epsilon^{-1}(\mathbf{G},\mathbf{G'})$) computed neglecting the non-diagonal components for a prototypical semiconductor AlP, and a prototypical insulator MgO. 
We used two model functions to fit  $\epsilon^{-1}(\mathbf{G},\mathbf{G'})$ and extract screening parameters that we collectively refer to as $\mu_{\text{PDEP}}$\footnote{PDEP here refers to the projective dielectric eigendecomposition technique used to evaluate the dielectric screening}: the Thomas-Fermi screening model (corresponding to a Yukawa potential) and the complementary error function. Using the former, the screened Coulomb interaction is:
\begin{equation}
W(\mathbf{r},\mathbf{r'}) \approx
\frac{\epsilon_\infty^{-1}}{|\mathbf{r}-\mathbf{r'}|} +  ( 1 - \epsilon_{\infty}^{-1} ) \frac{e^{-k_{TF} \mathbf{|r-r'|}}}{|\mathbf{r-r'}|} 
\label{eq:TFmodel}
\end{equation}
The fit of the dielectric function was carried out in Fourier space: 
\begin{equation}
\epsilon^{-1}_{TF}(\mathbf{G}) = \epsilon_\infty^{-1} + ( 1 - \epsilon_{\infty}^{-1} ) \frac{G^2}{G^2+k_{TF}^2} .
\label{eq:TFQ}
\end{equation}
If instead of the TF screening model, the complementary error function is used to fit $\epsilon^{-1}(\mathbf{G},\mathbf{G'})$, we obtain:
\begin{equation}
\epsilon^{-1}_{erfc}(\mathbf{G}) = \epsilon_\infty^{-1} + ( 1 - \epsilon_\infty^{-1} ) (1-e^{-\frac{G^2}{4 \mu^2}}) .
\label{eq:RSHQ}
\end{equation}

The performance of the RS DDH functional obtained by substituting in Eq.~\eqref{equation:GKSxc} $\alpha=\epsilon_{\infty}^{-1}$, $\beta=0.25$ and for each of the three $\mu$ parameters described above (namely $\mu_{\text{WS}}$, $\mu_{\text{TF}}$  and $\mu_{\text{PDEP}}$) will be discussed in Section III A for inorganic semiconductors and insulators; and in Section III B for molecular crystals. We will show that our results are largely insensitive to the three choices of $\mu$, which turn out to be similar to each other for all systems examined here. 

\subsection{Computational details}

The evaluation of the electronic dielectric constant was carried out within an all-electron approach using the coupled perturbed Kohn-Sham (CPKS)\cite{Rerat:2008th,Johnson:1993db} equations (the coupled-perturbed Hartree-Fock method (CPHF)\cite{Pople:1979hm,Hurst:1988gp,Orlando:2009fu} extended to DFT) as implemented in the CRYSTAL14\cite{Dovesi:2014ej} electronic structure package, where we explicitly computed $\epsilon$ beyond the random phase approximation (RPA) by evaluating $f_{xc} = \frac{\delta v_{xc}}{\delta \rho}$, i.e. the functional derivative of the nonlocal potential $v_{xc}$. The effect of including $f_{xc}$ in the evaluation of $\epsilon$ was discussed in Ref.~\onlinecite{Skone:2014hu}. The dielectric constant determined self-consistently was then used in the range-separated functional of Eq.~\eqref{eq:wstatRSH} as implemented in CRYSTAL14 and a development version\footnote{We implemented the general form of the range-separated hybrid described in the manuscript in a development version of Quantum-Espresso.}of the Quantum-ESPRESSO package.\cite{Giannozzi:2009hx} We note that a denser {\it k}-point mesh is required for the convergence of the electronic dielectric constants than for the KS eigenvalues (see Supplemental Material for details). 

The WEST\cite{Govoni:2015gc} code was used to compute the dielectric matrix using 1024 eigenpotentials for each system considered here, unless otherwise noted. These calculations were carried out at the $\Gamma$-point for a supercell of appropriate size.

For the all-electron calculations, we used Gaussian basis sets, modified starting from Alhrich's def2-TZVPP molecular basis,\cite{Weigend:2005dh} with the only exception of rare gases Ne and Ar basis sets, which were modified starting from the def2-QZVPD set.\cite{Rappoport:2010} For Co and Ni we used  the def2-TZVP modified basis sets of Bredow \textit{et al.}\cite{Peintinger:2012ie} For the plane-wave calculations, we adopted  norm-conserving pseudopotentials of the Troullier-Martins type,\cite{Troullier:1991wi} where for the transition metal atoms, unless otherwise noted, the $(n-1)$s and $(n-1)$p electrons were included in the valence ($n$ is the highest principal quantum number). Plane-wave kinetic energy cutoffs and localized Gaussian basis sets employed here can be found in the Supplemental Material. All calculations were performed at the experimental geometry and $T=0$K, without consideration of zero-point vibrational effects. 
In the case of transition metals we excluded any semi-core ($n - 1$)s and ($n - 1$)p electrons from the definition of the valence electron density that is used to determine $\mu_{\text{WS}}$ and $\mu_{\text{TF}}$.

To obtain the ionization potential of molecular crystals on an absolute scale (with respect to vacuum), vacuum slab model calculations were carried out using a plane-wave pseudopotential basis set, aligning the electrostatic potential of the bulk with that of the vacuum for each level of theory used. The supercell size in the direction perpendicular to the interface was chosen so as to ensure a converged value of the electrostatic potential (see the Supplemental Material for details).

In our calculation of gas phase ionization potentials discussed in Section III C, a plane-wave basis was used along with a Makov-Payne\cite{Makov:1995vk} correction to properly align the orbital eigenvalues with the vacuum level position. Molecular polarizabilities were evaluated at the PBE level of theory. Although the values of the polarizability varies with respect to the level of theory applied, the inverse cube root used to define the inverse molecular polarizability radius exhibits a negligible variation with respect to the level of theory. The screening parameters determined from the optimally tuned range-separted hybrid (OT-RSH) procedure described in Ref.~\onlinecite{RefaelyAbramson:2012uw} were evaluated in a planewave basis with a Makov-Payne correction applied, using the Quantum-ESPRESSO package.

\section{ \textbf {Results and Discussion}}
\label{results}
\subsection{{\bf Inorganic semiconductors and insulators }}
\begin{figure}[]
\centering
\scalebox{0.32}{\includegraphics[trim= 5mm 5mm 70mm 30mm,clip, angle=0]{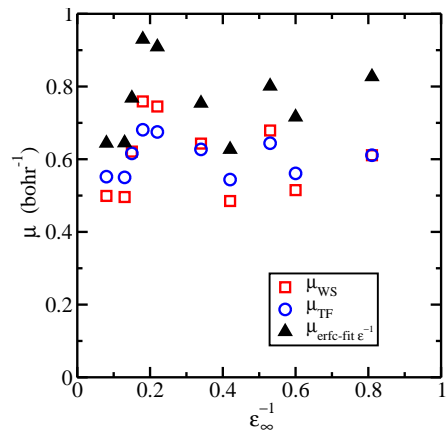}}
\caption{(Color online) Screening parameters $\mu$ (as defined in the text) are plotted versus the inverse electronic dielectric constant for a set of inorganic semiconductors and insulators. The Wigner-Seitz screening parameter $\mu_{WS}$ (red open squares) Eq.~\eqref{eq:WS}, the Thomas-Fermi screening parameter $\mu_{\text{TF}}$ (blue open circles) Eq.~\eqref{eq:TFsimplify}, and $\mu_{\text{erfc-fit}}$ (black filled triangles) obtained from a fit to Eq.~\eqref{eq:RSHQ}, are shown. }
\label{fig:plot-screen-inorganic}
\end{figure}

The non empirical values of the screening parameters for a diverse set of inorganic semiconductors and insulators as obtained from Eq.~\eqref{eq:WS} and Eq.~\eqref{eq:TFsimplify} are provided in Table~\ref{table:mu-tab}. For a subset of systems the screening parameters were also computed by fitting the long-range decay of the dielectric function to either Eq.~\ref{eq:TFQ} or to Eq.~\ref{eq:RSHQ}, and these are listed under the column heading $\mu_{\text{PDEP}}$.
The parameters $\mu_{\text{PDEP}}$ are in general larger than $\mu_{\text{WS}}$ and $\mu_{\text{TF}}$, which are similar to each other. However as we will see below, functionals defined with each of these three choices yield very similar results for the electronic gaps. A graphical comparison of screening parameters as a function of the dielectric constant is given in Fig.~\ref{fig:plot-screen-inorganic}.
 
\begin{table}[] 
\centering
\caption{The dielectric constant ($\epsilon_{\infty}$) determined self-consistently as described in Ref.~\onlinecite{Skone:2014hu} for the set of semiconductors and insulators listed in the first column, is given in column 2. The screening parameters ($\mu$) used in the RSH functional form of Eq.~\eqref{eq:wstatRSH} are listed in units of bohr$^{-1}$ in columns 3-6. (See text) }
\label{table:mu-tab}
\setlength{\tabcolsep}{4pt}
\begin{tabular}{lccccc}
\hline
\hline
 &                         &                  &                                               &  \multicolumn{2}{c}{$\mu_{\text{PDEP}}$} \\ 
 &                         &                  &                                               &  \multicolumn{2}{c}{$\overbrace{\rule{2.2cm}{0cm}}$} \\[-5pt] 
 & $\epsilon_{\infty}$ &   $ \mu_{\text{WS}}$ & $ \mu_{\text{TF}}$ & $\mu_{\text{erfc-fit}}$ & $\mu_{\text{TF-fit}}$    \\  
\hline
  Si          & 11.76&   0.50  & 0.55     & 0.64 & 0.64\\
  AlP         & 7.23 &   0.50  & 0.55     & 0.65 & 0.64\\
  SiC         & 6.50 &   0.62  & 0.62     & 0.77 & 0.77\\ 
  TiO$_2$     & 6.56 &   0.68  & 0.65     &      &     \\
  NiO         & 5.49 &   0.82  & 0.71     &      &     \\
  C           & 5.61 &   0.76  & 0.68     & 0.93 & 0.97\\
  CoO         & 4.92 &   0.78  & 0.69     &      &     \\
  GaN         & 5.14 &   0.60  & 0.61     &      &     \\
  ZnS         & 4.95 &   0.65  & 0.63     &      &     \\
  MnO         & 4.45 &   0.72  & 0.66     &      &     \\
  WO$_3$      & 4.72 &   0.66  & 0.63     &      &     \\
  BN          & 4.40 &   0.75  & 0.68     & 0.91 & 0.95\\
  HfO$_2$     & 3.97 &   0.66  & 0.63     &      &     \\
  AlN         & 4.16 &   0.49  & 0.55     &      &     \\
  ZnO         & 3.46 &   0.78  & 0.69     &      &     \\
  Al$_2$O$_3$ & 3.01 &   0.71  & 0.66     &      &     \\
  MgO         & 2.81 &   0.64  & 0.63     & 0.75 & 0.72 \\
  LiCl        & 2.77 &   0.53  & 0.57     &      &     \\
  NaCl        & 2.29 &   0.49  & 0.54     & 0.63 & 0.64\\
  LiF         & 1.86 &   0.68  & 0.64     & 0.80 & 0.83\\
  H$_2$O      & 1.68 &   0.55  & 0.58     & 0.52 & 0.53\\
  Ar          & 1.66 &   0.52  & 0.56     & 0.72 & 0.73\\
  Ne          & 1.21 &   0.61  & 0.61     & 0.83 & 0.89\\
\hline
\hline
\end{tabular}
\end{table}

Photoemission gaps of inorganic semiconductors and insulators\footnote{For the present analysis, Ge was removed from the original test set because of uncertainties in the accuracy of the localized basis sets and pseudopotentials.} 
computed with the $\mu$ parameters of Table~\ref{table:mu-tab} are reported in Table~\ref{table:Eg-comp-tab}.
For most systems, irrespective of the choice of the system dependent screening parameter $\mu$, the RSH functional yields improved electronic gaps over the already accurate full-range sc-hybrid functional. The only exceptions, are some transition metal oxides where partially occupied d-orbitals dominate the character of the valence band--most notably CoO as well as FeO (not shown) and to a lesser extent NiO and MnO\footnote{Though the occupied valence bands of NiO and MnO have partial d-character, they also show a stronger hybridization with the oxygen 2p orbitals than the valence bands of CoO and FeO, and hence they are less localized. See Supplemental Material for further discussion}. 
If these systems are not included in the assessment of the RSH functional quality, a more noticeable improvement is obtained, as shown in Table S3, and by comparison with self-consistent GW results from Shishkin \textit{et al.}\cite{Shishkin:2007PRL} (see column 7 of Table~\ref{table:Eg-comp-tab}). As shown in Table S3, for this subset of solids the MARE for all definitions of $\mu$ are lower than that of the sc-GW results.

\begin{table*}[] 
\caption{The Kohn-Sham (KS) energy gaps (eV) evaluated with hybrid functionals are compared with the experimental (Exp.) electronic gaps for a wide range of semiconductors and insulators. The experimental values correspond to either photoemission measurements or to optical measurements where the excitonic contributions were removed, with alumina being the only exception. The KS gaps were computed as the single particle energy difference of the conduction band minimum and the valence band maximum. The sc-hybrid  heading refers to hybrid calculations where the fraction of exact-exchange is the self-consistent ${\epsilon}_{\infty}$. The RSH columns correspond to the electronic gap evaluated with the range separation scheme described in Section II, and the screening parameters $\mu_{\text{WS}}$ (Eq.~\ref{eq:WS}), $\mu_{\text{TF}}$ (Eq.~\ref{eq:TFsimplify}), and $\mu_{\text{erfc-fit}}$ (Eq.~\ref{eq:RSHQ}). The self-consistent quasiparticle (scGW) gaps are taken from Ref.~\citenum{Shishkin:2007PRL}.
ME, MAE, MRE, and MARE are the mean, mean absolute, mean relative, and mean absolute relative error, respectively. The experimental geometry was used in all calculations. Note that CoO, NiO, and MnO are magnetic with AFM-II magnetic ordering. The structure/polytype used for each system is the same as in Table I of Ref.~\onlinecite{Skone:2014hu}.} 
\label{table:Eg-comp-tab}
\setlength{\tabcolsep}{8pt}
\begin{tabular}{lccccccc}
\hline
\hline
           & PBE0 &  sc-hybrid& RSH                &  RSH               & RSH                      & sc$GW$\cite{Shishkin:2007PRL} & Exp.  \\  
           &      &           &  $\mu_{\text{WS}}$ & $\mu_{\text{TF}}$  & $\mu_{\text{erfc-fit}}$  &   &  \\
\hline
  Si           & 1.75  &  0.99 & 1.03 & 1.02 &   1.01 &1.24& 1.17 \cite{Kittel:2005} \\
  AlP          & 2.98  &  2.37 & 2.43 & 2.42 &   2.40 &2.57& 2.51 \cite{Monemar:1973jj} \\
  SiC          & 2.91  &  2.29 & 2.32 & 2.32 &   2.31 &2.53& 2.39 \cite{Choyke:1964fh} \\ 
  TiO$_2$      & 3.92  &  3.05 & 3.16 & 3.17 &        &    & 3.3  \cite{Tezuka:1994cv} \\
  NiO          & 5.28  &  4.11 & 4.45 & 4.51 &        &    & 4.3  \cite{Sawatzky:1984jt} \\
  C            & 5.95  &  5.42 & 5.44 & 5.45 &   5.43 &5.79& 5.48 \cite{Clark:1964km} \footnote{The exp. QP gap reported here does not account for the zero-point vibrational gap renormalization, which has been shown to be nonnegligible for diamond.\cite{Cardona:2005ho,Giustino:2010ek,Antonius:2014im}} \\
  CoO          & 4.53  &  3.62 & 3.92 & 3.98 &        &    & 2.5  \cite{vanElp:1991ia} \\
  GaN          & 3.68  &  3.26 & 3.30 & 3.30 &        &3.27& 3.29 \cite{RamirezFlores:1994ws}\\
  ZnS          & 4.18  &  3.82 & 3.85 & 3.86 &        &3.60& 3.91 \cite{Kittel:2005} \\
  MnO          & 3.87  &  3.60 & 3.65 & 3.49 &        &    & 3.9 \cite{Meyer:2010cr} \\
  WO$_3$       & 3.76  &  3.47 & 3.49 & 3.49 &        &    & 3.38 \cite{Meyer:2010cr} \\
  BN           & 6.51  &  6.33 & 6.33 & 6.34 &   6.33 &6.59& 6.4 \cite{Levinshtein:2001} \\ 
  HfO$_2$      & 6.65  &  6.68 & 6.67 & 6.67 &        &    & 5.84 \cite{Sayan:2004gx}  \\
  AlN          & 6.31  &  6.23 & 6.22 & 6.23 &        &    & 6.28 \cite{Roskovcova:1980gz} \\
  ZnO          & 3.41  &  3.78 & 3.75 & 3.67 &        &3.2 & 3.44 \cite{Ozgur:2005it} \\
  Al$_2$O$_3$  & 8.84  &  9.71 & 9.63 & 9.61 &        &    & 8.8  \cite{Innocenzi:1990bx} \\
  MgO          & 7.25  &  8.33 & 8.23 & 8.22 &   8.27 &8.12& 7.83 \cite{Whited:1973eo} \\
  LiCl         & 8.66  &  9.62 & 9.52 & 9.54 &        &    & 9.4  \cite{Baldini:1970ba} \\
  NaCl         & 7.26  &  8.84 & 8.60 & 8.66 &   8.71 &    & 8.6  \cite{Nakai:1969wu} \\
  LiF          & 12.18 & 15.69 &15.24 & 15.18&15.42 &14.5& 14.2  \cite{Piacentini:1976fw} \\
  H$_2$O       & 7.92  & 11.49 &10.89 & 10.94&10.84 &    & 10.9  \cite{Kobayashi:1983wx} \\
  Ar           & 11.20 & 14.67 &14.12 & 14.20&14.41 &13.9& 14.2  \cite{Schwentner:1975dp} \\
  Ne           & 15.20 & 23.67 &21.44 & 21.44&22.28 &21.4& 21.7  \cite{Schwentner:1975dp} \\
\hline
  ME (eV)   &-0.40  & 0.32 & 0.18 & 0.19 &  &  & -- \\
  MAE (eV)  & 1.08  & 0.42 & 0.29 & 0.30 &  &  & -- \\
  MRE (\%)  &  6.2  & 3.7  & 3.4 & 3.5   &  &  & -- \\
  MARE (\%) & 17.1  & 7.5  & 6.4 & 6.5   &  &  & -- \\
\hline
\hline
\end{tabular}
\end{table*}

In the case of $\mu_{\text{PDEP}}$, we only tabulate the computed electronic gaps for the complementary error function fit, using Eq.~\eqref{eq:RSHQ}, labeled as $\mu_{\text{erfc-fit}}$ in Table~\ref{table:Eg-comp-tab} since those  obtained from fitting to the Thomas-Fermi screening model (Eq.~\eqref{eq:TFsimplify}) are nearly identical. Overall the RSH results using $\mu_{\text{erfc-fit}}$ appear to be very similar to those of the sc-hybrid functional.

The small differences between the MAEs of results obtained with the various choices of $\mu$ warrant further analysis to determine whether these apparent differences are statistically meaningful. We determined the confidence interval over which one methodology MAE is statistically different from another one by performing a Wilcoxon signed-rank test\cite{Wilcoxon:1945} between pairs of methods; a p-value below 0.05\cite{Bauer:1972} was chosen to indicate that indeed the two sets of results exhibit different MAEs (i.e. differences are not due to chance). We found that the p-values computed between the MAE  obtained with different choices of $\mu$ are all larger than 0.05, indicating that the three choices of $\mu$ yield the same result. For any of the RS-DDH functionals compared to the global sc-hybrid or PBE0 functioinal, all p-values are below 0.05 indicating that the improvement in the MAE of RSH functionals relative to sc-hybrid or PBE0 is statistically significant. For a summary of all p-values between pairs of methods for the full set of 23 systems listed in Table~\ref{table:Eg-comp-tab} and for a subset of these solids where transition metal oxides are removed, see Table S5 and Table S6, respectively, in the Supplemental Material.      

\begin{figure}[h!]
 \centering
\begin{subfigure}{0.9\linewidth}
  \centering
  \scalebox{1.0}{ \includegraphics[width=1.0\textwidth,trim=15mm 0mm 15mm 05mm, clip]{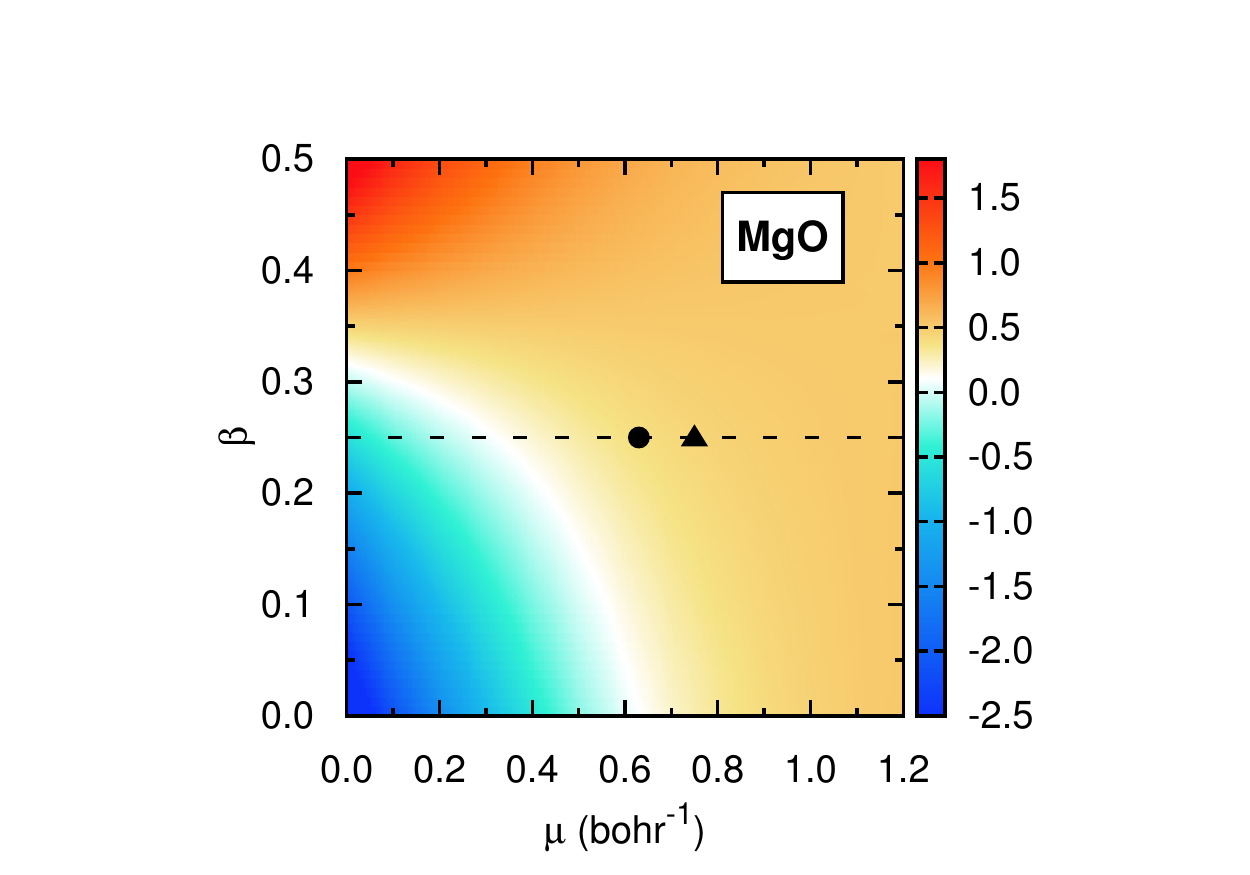}}\hfill 
\end{subfigure}%
\\
\begin{subfigure}{1.0\linewidth}
  \centering
  \scalebox{0.9}{ \includegraphics[width=1.0\textwidth, trim=15mm 0mm 15mm 05mm, clip]{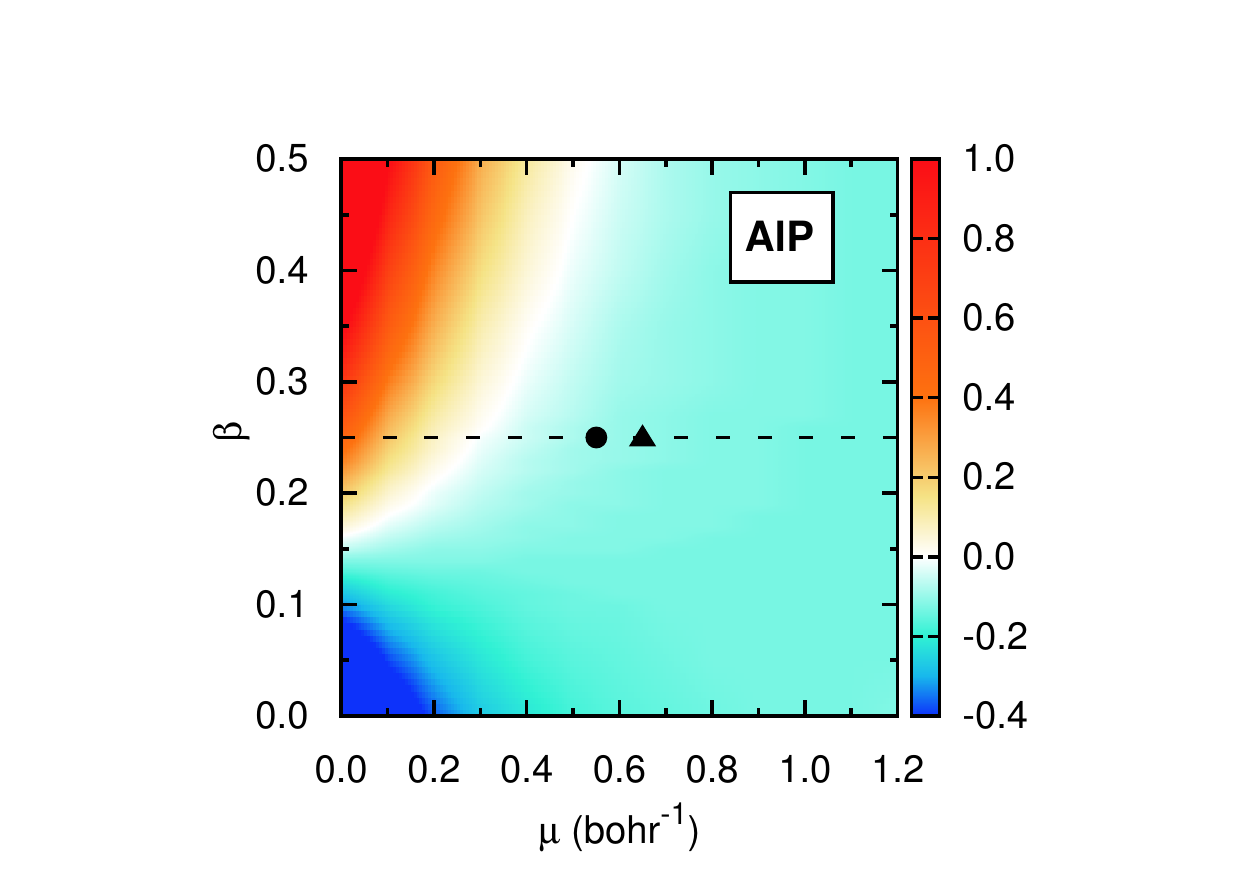}} 
\end{subfigure}
\caption{(Color online) Signed electronic gap errors (theory - experiment), in units of eV, for a RSH functional defined by Eq.~\eqref{eq:wstatRSH}, as a function of two parameters: $\mu$ and $\beta$ (see text). Results are reported for a prototypical insulator (MgO) with $\epsilon_{\infty} < 4$ and a semiconductor (AlP) with $\epsilon_{\infty} > 4$. Positive (negative) errors are indicated by shades of red (blue). The white shaded area corresponds to zero error with respect to experiment. 
The dashed black line corresponds to the fixed parameter space explored in the RS-DDH form where $\beta = 0.25$. The black circle and triangle indicate where the values of $\mu_{\text{TF}}$ and $\mu_{\text{erfc-fit}}$ fall, respectively.}
\label{fig:MgO-AlP-2D-plot}
\end{figure}

To assess how the electronic gap depends on the sr fraction of exchange we examined the gap dependence for two systems with $\epsilon > 4$ and $\epsilon < 4$, respectively. In Fig.~\ref{fig:MgO-AlP-2D-plot} we set $\alpha=\epsilon^{-1}_{\infty}$ in the RSH functional form and we show the behavior of the signed electronic gap error as a function of $\mu$ and $\beta$.
The prototypical semiconductor (AlP) and insulator (MgO) show opposite behaviors and the curvature of the minimum error (white region) is concave for the former and convex for the latter. As $\mu \rightarrow \infty $ the sc-hybrid functional is recovered and an overestimation (underestimation) of the gap for insulating MgO (semiconducting AlP) is obtained.   

%
\subsection{{\bf Molecular Crystals }}

\begin{table}[] 
\caption{The dielectric constant ($\epsilon_{\infty}$) determined self-consistently as described in Ref.~\onlinecite{Skone:2014hu} for the set of molecular crystals listed in the first column, is given in column 2. The screening parameters ($\mu$) used in the RSH functional form of Eq.~\eqref{eq:wstatRSH} are listed in units of bohr$^{-1}$ in columns 3-4. All $\mu$ have units of bohr$^{-1}$.}
\label{table:mu-tab-molxtal}
\setlength{\tabcolsep}{4pt}
\begin{tabular}{lccc}
\hline
\hline
 & & \multicolumn{2}{c}{Crystalline} \\
 &  $\epsilon_{\infty}$ &    $ \mu_{\text{TF}}$ &  $\mu_{\text{erfc-fit}}$ \\
\hline
  C$_{14}$H$_8$S$_4$-C$_{12}$H$_4$N$_4$ (DBTTF-TCNQ)   &11.07 & 0.58 &      \\
  C$_{60}$ (buckminsterfullerene)                      & 4.29 & 0.61 &  0.57\\
  C$_{32}$H$_{18}$N$_8$ (phthalocyanine)               & 3.97 & 0.60 &      \\
  C$_{24}$H$_{8}$O$_6$ ($\alpha$ PTCDA)                & 3.45 & 0.61 &       \\ 
  C$_{22}$H$_{14}$ (pentacene)                         & 3.36 & 0.59 &       \\
  C$_{20}$H$_{12}$O$_2$ ($\beta$ quinacridone)         & 3.15 & 0.60 &       \\
  C$_{18}$H$_{12}$ (tetracene)                         & 3.15 & 0.59 &       \\
  C$_{14}$H$_{10}$ (anthracene)                        & 3.02 & 0.58 &       \\
  C$_{42}$H$_{28}$ (rubrene)                           & 2.88 & 0.58 &  0.50 \\
  C$_{10}$H$_8$ (naphthalene)                          & 2.70 & 0.58 &       \\
  C$_6$H$_6$ (benzene)                                 & 2.40 & 0.57 &  0.54 \\
  NH$_3$ (ammonia)                                     & 2.00 & 0.57 &  0.53 \\
  C$_2$H$_4$O$_2$ (acetic acid)                        & 1.88 & 0.60 &       \\
  H$_2$O (ice)                                         & 1.68 & 0.58 &  0.52 \\
\hline
\hline
\end{tabular}
\end{table}

\begin{table*}[]
\caption{The $xx$, $yy$, and $zz$ componenets of the dielectric tensor ($\epsilon_{\infty}$) of the systems listed on the first row, computed at the PBE, PBE0, and sc-hybrid levels of theory and the corresponding experimental results.}
\label{table:epsilinf-anisotropic}
\begin{tabular}{lccccccccccc}
\hline
\hline
 & \multicolumn{3}{c}{Anthracene}& &\multicolumn{3}{c}{PTCDA} & &\multicolumn{3}{c}{DBTTF-TCNQ}   \\
 \hline
               &   $\epsilon_{xx}$  &  $\epsilon_{yy}$ &  $\epsilon_{zz}$ &&   $\epsilon_{xx}$  &  $\epsilon_{yy}$  &  $\epsilon_{zz}$&&   $\epsilon_{xx}$  &  $\epsilon_{yy}$ &  $\epsilon_{zz}$ \\
\hline
PBE         & 2.28 & 2.93 & 4.30  && 2.25  &4.40  & 4.50  && 42.38  & 3.67 & 2.90 \\ 
PBE0        & 2.22 & 2.83 & 4.10  && 2.12  &4.10  & 4.12  && 11.08  & 5.82 & 2.72 \\ 
sc-hybrid   & 2.21 & 2.80 & 4.02  && 2.11  &4.06  & 4.08  && 26.28  & 3.54 & 2.81  \\ 
Exp.\cite{Cummins:1974vv, Alonso:2002vr} & 2.42 $\pm$0.05 & 2.90 $\pm$0.05  & 4.07 $\pm$0.05 && 2.40  & 5.29  & 5.02  &&  &  &  \\ 
Exp.\cite{Munn:1973ba, Zang:1991gq}        & 2.62 $\pm$0.03 & 2.94 $\pm$0.03  & 4.08 $\pm$0.03 && 1.90  &  & 4.49  &&  &  &  \\ 
Exp.\cite{Karl:1971hm, Fuchigami:1995jp}   & 2.51  & 2.99   &  4.11 && 2.28  &  & 3.73  &&  &  &  \\ 
\hline
\hline
\end{tabular}
\end{table*}
 
\begin{figure}[]
\centering
\scalebox{0.32}{\includegraphics[trim= 5mm 5mm 50mm 30mm,clip, angle=0]{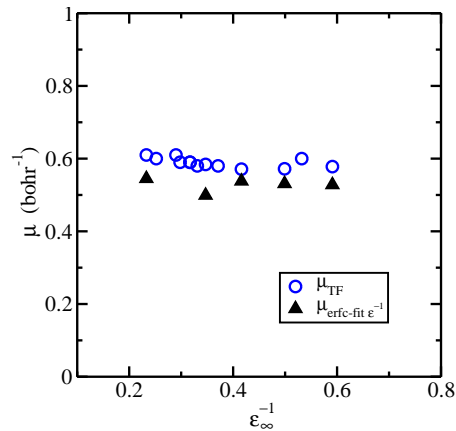}}
\caption{(Color online) Several values of screening parameter $\mu$ (see text) plotted as a function of the inverse electronic dielectric constant ($\epsilon^{-1}_{\infty}$) for the set of molecular crystals of Table~\ref{table:mu-tab-molxtal}. See Eq.~\eqref{eq:TFsimplify} for $\mu_{\text{TF}}$ and Eq.~\eqref{eq:RSHQ} for $\mu_{\text{erfc-fit}}$.  }
\label{fig:plot-screen-molec}
\end{figure}

The dielectric constants determined self-consistently for a set of molecular crystals are shown in Table~\ref{table:mu-tab-molxtal} along with the screening parameter $\mu_{\text{TF}}$, which is found to be nearly constant at $~0.58$ bohr$^{-1}$ and $\mu_{\text{erfc-fit}}$, which is similar albeit slightly smaller than $\mu_{\text{TF}}$. 
The screening parameters $\mu$ are shown as a function of $\epsilon^{-1}_{\infty}$ in Fig.~\ref{fig:plot-screen-molec}. 
Comparison between different levels of theory for the computed dielectric constants are shown in Table S2 of the Supplemental Material. 
In general, the PBE results yield the poorest agreement with experiment, with PBE0 showing a marked improvement, and the DDH functionals performing the best. 

For a subset of optically anisotropic molecular crystals, we compare in Table~\ref{table:epsilinf-anisotropic} the dielectric tensor components computed at different levels of theory with that obtained experimentally. As expected, the values of the tensor components vary depending on the level of theory, but they all exhibit similar errors with respect to experiment. At variance with inorganic materials,\cite{Skone:2014hu} the dielectric constant of organic crystals shows a weaker dependence on the functional used. A notable exception is the charge-transfer molecular crystal DBTTF-TCNQ, where along the charge transfer direction, $\epsilon_{xx}(\infty)$ varies from 42.38 (PBE) to 11.08 (PBE0). Clearly for charge-transfer molecular crystals the use of a self-consistently determined dielectric screening and the corresponding set of wavefunctions are required to obtain accurate dielectric constants.\footnote{In the present RS DDH the dielectric constant and consequently the fraction of excahnge is an average over all directions. Computing quantities such as charge transfer integrals\cite{Fonari:2015dx} would likely require a more explicit handling of the directional screening, e.g. by evaluating the full dielectric screening matrix $W$ in $GW$ based methods.} \\ 

Below we discuss results obtained for electronic gaps, and vertical ionization potentials obtained using the parameters reported in Table~\ref{table:mu-tab-molxtal}.

The computed electronic gaps are compared with experiment in Table~\ref{table:Eg-comp-molec-cry}. In general all the DDH functionals give similar results; the reason why there is almost no variability  between the full-range and range-separated hybrids for the organic molecular crystals stems from their inverse dielectric constant (0.2-0.35) which is close to the fraction of exchange adopted for the short-range limit (0.25), making the lr and sr limits of RSH very similar. To investigate the statistical significance of our results, we performed a paired test between the MAEs of the DDH functionals, showing that the MAEs of the full-range and the range-separated DDH functionals are not statistically significant (see Table S7 of Supplemental Material), unlike the case of inorganic materials. 

In a previous study,\cite{Fonari:2014cy} it was observed that a fraction of exchange in the range of 0.31-0.35 yielded electronic gaps of organic semiconducting molecular crystals, such as pentacene, in very good agreement with experiment and GW calculations. Indeed, as mentioned above and shown in Table S2, the computed inverse dielectric constants of many of the semiconducting molecular crystals are centered around $\sim$0.3.  We also note that the good performance of the PBE0-1/3 functional\cite{Guido:2013iv} for molecular crystals reported in the literature most likely stems from the 1/3 fraction of exact-exchange used being similar in value to the inverse dielectric constant of the molecular crystals.

\begin{figure}[]
\centering
\scalebox{0.80}{\includegraphics[trim= 15mm 00mm 15mm 05mm,clip, angle=0]{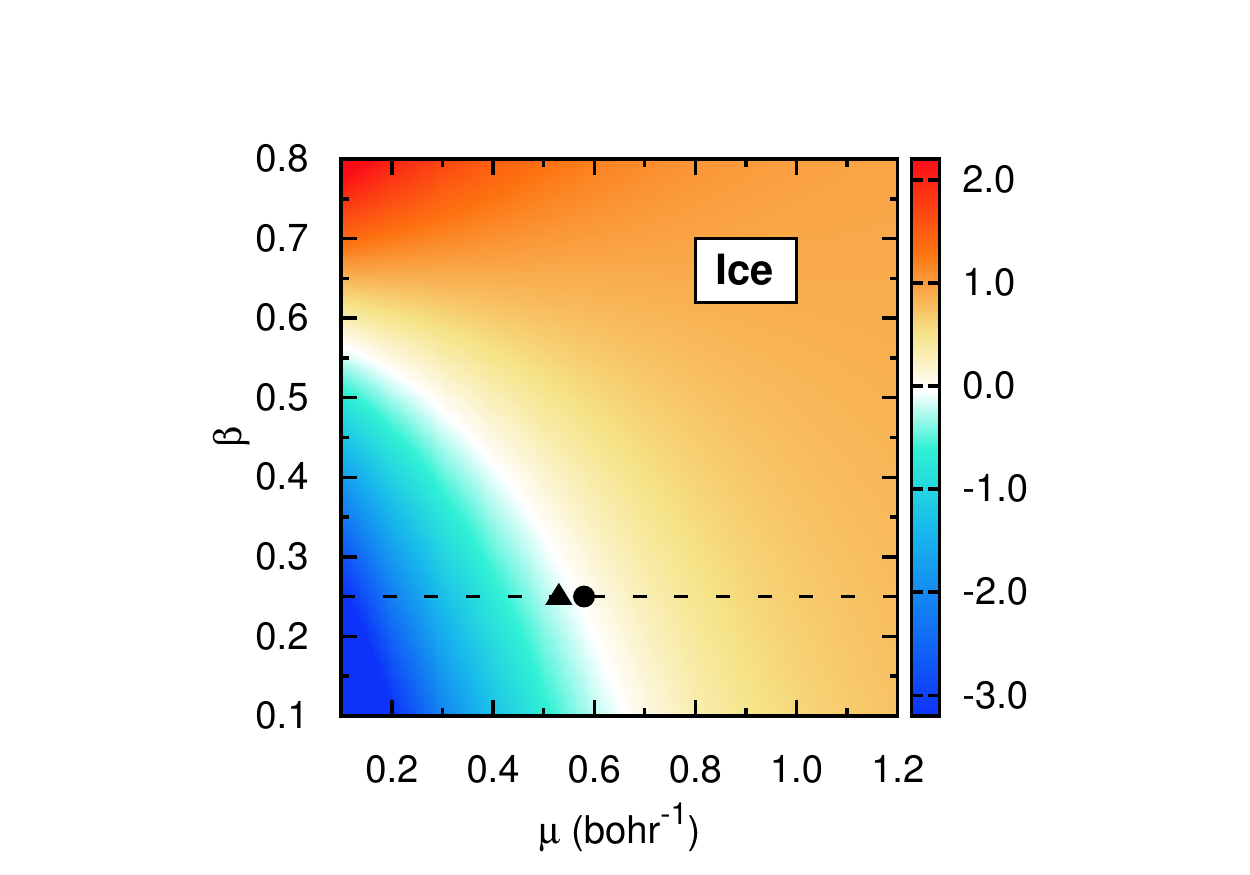}}
\caption{(Color online) Signed electronic gap error of ice (theory - experiment), in units of eV, for a RSH functional defined by Eq.~\eqref{eq:wstatRSH}. Positive errors are indicated by shades of red colors and negative errors by shades of blue color. The white shaded area corresponds to zero error of the RSH with respect to experiment. The black dashed line is the short-range value of exact exchange for $\beta =0.25$. The black circle and triangle indicate where the values of $\mu=\mu_{\text{TF}}$ and $\mu_{\text{erfc-fit}}$ fall, respectively. }
\label{fig:ice-plot}
\end{figure}
Similar to what was done for two exemplary inorganic systems, we varied the sr fraction of Hartree-Fock exchange ($\beta$) together with the screening parameter ($\mu$) for a molecular crystal (ice), to assess how the electronic gap depends on the coupling of these two quantities. Fig.~\ref{fig:ice-plot} shows the 2D heat map of the signed electronic gap error for ice as a function of $\beta$ and $\mu$. A concave curvature of the minimum error (white region) similar to that of the inorganic insulator MgO with $\epsilon_{\infty} < 4$ was observed for ice. 

\begin{table*}[] 
\caption{The Kohn-Sham (KS) energy gaps (eV) evaluated with the dielectric-dependent hybrid functionals, PBE and PBE0 are compared with the experimental electronic gaps for several molecular crystals. The experimental values are from photoemission measurements. }
\footnotetext[1]{See the Supplemental Material for details on the cell of ice used. The experiemtnal photoemmision gap shown is for proton-disordered ice Ih @ 80K. }
\label{table:Eg-comp-molec-cry}
\begin{tabular}{lccccccc}
\hline
\hline
               &  PBE & PBE0 & Hybrid & Hybrid   & sc-hybrid&  RSH & Exp.  \\  
&  $\alpha=0$  & $\alpha=0.25$   & $\alpha={1/\epsilon}^{\text{PBE}}_{{\infty} }$ & $\alpha={1/\epsilon}^{\text{PBE0}}_{{\infty} }$ & $\alpha={1/\epsilon}^{\text{sc}}_{{\infty} }$& $\mu_{\text{TF}}$ & \\  
\hline
  DBTTF-TCNQ                  & 0.16 & 0.74 & 0.26  & 0.45 & 0.31  & 0.33 &  \cite{}\\
  C60                         & 1.27 & 2.34 & 2.11  & 2.27 & 2.26  & 2.26 & 2.3 $\pm$ 0.1 \cite{Lof:1992uu}\\
  phthalocyanine (H$_2$Pc)    & 1.22 & 1.85 & 1.84  & 1.85 & 1.85  & 1.85 & 2.2 $\pm$ 0.2 \cite{Zahn:2006hj}\\
  PTCDA                       & 1.41 & 2.53 & 2.62  & 2.72 & 2.73  & 2.73 & 2.74 $\pm$ 0.2 \cite{Zahn:2006hj}\\ 
  pentacene                   & 0.76 & 1.83 & 1.95  & 2.04 & 2.05  & 2.05 & 2.1 \cite{Sato:1981dn, Lindstrom:2007bb}\\
  quinacridone                & 1.43 & 2.76 & 3.02  & 3.13 & 3.13  & 3.11 &   \cite{}\\
  tetracene                   & 1.26 & 2.46 & 2.72  & 2.78 & 2.80  & 2.79 & 3.3 \cite{Sato:1981dn, Lindstrom:2007bb}\\
  anthracene                  & 2.05 & 3.45 & 3.82  & 3.89 & 3.91  & 3.89 & 3.72 \cite{Vaubel:1968ve}\\
  rubrene                     & 1.15 & 2.32 & 2.70  & 2.77 & 2.79  & 2.77 & 2.67 \cite{Ding:2009cu}\\
  naphthalene                 & 3.05 & 4.64 & 5.32  & 5.39 & 5.41  & 5.37 & 5.29 \cite{Sato:1981dn,Lindstrom:2007bb}\\
  benzene                     & 4.57 & 6.37 & 7.48  & 7.56 & 7.58  & 7.47 & 7.58 \cite{Sato:1981dn, Lindstrom:2007bb}\\
  ammonia                     & 4.52 & 6.82 & 8.76  & 9.00 & 9.17  & 8.92 &  \cite{}\\
  acetic acid                 & 5.18 & 7.96 & 10.76 &10.95 &11.12  &10.62 &  \cite{}\\
  H$_2$O (ice\footnotemark[1])& 5.42 & 7.92 & 10.96 &11.23 &11.49  &10.94 & 10.9  \cite{Kobayashi:1983wx}  \\
\hline
  ME (eV)   &-2.05 &-0.69  &-0.12 &-0.02 &0.02  & -0.06 &  -- \\
  MAE (eV)  & 2.05 & 0.70  & 0.19 & 0.18 &0.21  & 0.16  &  --   \\
  MRE (\%)  &-49.6 &-13.5  &-4.9  &-2.3  &-1.7  &-2.6   &   -- \\
  MARE (\%) & 49.6 &13.8   &6.1   &5.1   & 5.4  &4.9    &   --   \\
\hline
\hline \end{tabular}
\end{table*}

We also evaluated the vertical ionization potential (vIP) of a subset of molecular crystals, namely rubrene, ice, and benzene, which are tabulated in Table~\ref{table:absoluteIP-solid}. Slab calculations were used (see Method Section) in order to place the valence band obtained from plane-wave pseudopotential calculations on an absolute energy scale (see Table S8 of the Supplemental Material for the electrostatic potential alignments compted at each level of theory). 
The results for PBE and PBE0 functionals did not provide very good agreement with experiment, whereas the sc-hybrid and RSH functional results appear to be in excellent agreement with experiment. However, we note that the experimental results in Table~\ref{table:absoluteIP-solid} may not correspond to the cleaved surface used in our calculation, though the rubrene surface explored here is a typically exposed surface of that molecular crystal when deposited on an oxide surface (e.g. indium tin oxide or silica).\cite{Chen:2014ks} In the case of ice, it appears that the sc-hybrid functional yields the best vertical ionization potential, but as shown by the gaps of Table~\ref{table:Eg-comp-molec-cry}, not necessarily the best gap. 

Finally we analyzed the photoelectron spectra of several molecular crystals, including ice, rubrene, and pentacene. We found that the position of the highest excitation with respect to vaccum is accurately described by both RS DDH and DDH functionals. The comparion of the full spectra with the corresponding experimental results will be given elsewhere.
\begin{table}[]
\caption{The vertical ionization potential (vIP$_s$), in units of eV, of several solid molecular crystals evaluated with PBE, PBE0, sc-hybrid and RSH functionals. The experimental values are listed for comparison. Note that for rubrene the surface listed corresponds to the orthorhombic cell. The screeneing parameters $\mu_{\text{TF}}$ and $\mu_{{\alpha_{\text{M}}}}$ are defined in Eq.~\eqref{eq:TFsimplify} and Eq.~\eqref{eq:molec_polariz}, respectively.} 
\footnotetext[1]{The prisim surface of ice is used. See Ref.~\onlinecite{Pan:2010ic} for further details on the common surfaces of ice.}
\label{table:absoluteIP-solid}
\begin{tabular}{lcccccc}
\hline
\hline
 &         surface &   PBE &  PBE0 & schybrid & RSH  &  Exp. \\
 &                 &       &       &          &$\mu_{\text{TF}}$ & \\
\hline
  rubrene   & (100)     & 3.85 & 4.45  & 4.69 & 4.68& 4.85\cite{Nakayama:2008ge}  \\
  benzene   & (001)    & 6.08 & 7.02  & 7.63 &7.61 &7.58\cite{Sato:1981dn} \\
 ice\footnotemark[1]       & (10$\overline{1}$0)    & 7.2 & 8.7  & 11.2 &10.7&11.8 \cite{Winter:2004vx, Nordlund:2001}\\
\hline
\hline
\end{tabular}
\end{table}

\subsection{ \textbf {Finite Systems: Molecules}}
Here we investigate the applicability of the RSH functional form to finite systems, including molecules and nanoparticles.  For finite systems $\epsilon_{\infty} \rightarrow 1$ and hence we discuss only the range-separated DDH. The generalization of a full-range DDH functional to finite systems is presented in Ref.~\onlinecite{Brawand:2016}.
In the case of molecules, we considered an additional screeneing parameter defined as:
\begin{equation}
\mu_{\alpha_{\text{M}}} = \left( \frac{1}{ \alpha_M} \right)^{\frac{1}{3}}
\label{eq:molec_polariz}
\end{equation} \\
where $\alpha_{\text{M}}$ is the molecular polarizability. 
We show below that a RSH form with $\mu = \mu_{\alpha_{{\text M}}}$ (Eq.~\eqref{eq:molec_polariz}) yields results in good agreement with experiments; however if one chooses $\mu_{\text{erfc-fit}}$ (which yields accurate results for both inorganic and organic solids), the agreement with experiments for isolated molecules is worsened. 

\begin{table}[] 
\caption{Screening parameters for isolated molecules. The second and third columns list the screening parameters obtained from the fit of the RPA dielectric function of the isolated molecules and obtained from the  molecular polarizability radius (see Eq.~\eqref{eq:molec_polariz}). We also give in column four the screening parameters obtained from the OT-RSH functional defined in Ref.~\onlinecite{RefaelyAbramson:2012uw}.} 
\label{table:mu-tab-otrsh}
\setlength{\tabcolsep}{4pt}
\begin{tabular}{lccc}
\hline
\hline
 &      $\mu_{\text{erfc-fit}}$ & $\mu_{\alpha_{\text {M}}}$ & $\mu_{\text{OT-RSH}}$ \\
\hline
  C$_{60}$ (buckminsterfullerene)                 & 0.64      &  0.12  & 0.14 \\
  C$_{24}$H$_{8}$O$_6$ ($\alpha$ PTCDA)           & 0.62      &  0.14  & 0.14 \\ 
  C$_{22}$H$_{14}$ (pentacene)                    & 0.56      &  0.14  & 0.15 \\
  C$_{20}$H$_{12}$O$_2$ ($\beta$ quinacridone)    & 0.59      &  0.15  & 0.15 \\
  C$_{18}$H$_{12}$ (tetracene)                    & 0.58      &  0.16  & 0.16 \\
  C$_{14}$H$_{10}$ (anthracene)                   & 0.59      &  0.18  & 0.18 \\
  C$_{42}$H$_{28}$ (rubrene)                      & 0.61      &  0.12  & 0.11 \\
  C$_{10}$H$_8$ (naphthalene)                     & 0.61      &  0.21  & 0.21 \\
  C$_6$H$_6$ (benzene)                            & 0.63      &  0.24  & 0.21 \\
  NH$_3$ (ammonia)                                & 0.63      &  0.40  & 0.33 \\
  C$_2$H$_4$O$_2$ (acetic acid)                   & 0.69      &  0.31  & 0.27 \\
  H$_2$O (water)                                  & 0.67      &  0.46  & 0.38 \\
\hline
\hline
\end{tabular}
\end{table}

Table~\ref{table:mu-tab-otrsh} compares the screening parameters $\mu_{\alpha_{\text {M}}}$ to $\mu_{\text{erfc-fit}}$ for molecules; we also compare to those obtained from the optimally tuned range-separated hybrid (OT-RSH)\cite{RefaelyAbramson:2012uw} procedure, where lr and sr fractions of exchange used in the OT-RSH are equal to the same limits used in the present RS-DDH functional (i.e. lr = 1 and sr = 0.25). Interestingly, $\mu_{\alpha_{\text {M}}}$ and $\mu_{_{\text{OT-RSH}}}$ are very similar for several molecular systems and are loosely dependent on molecular size, whereas $\mu_{\text{erfc-fit}}$ is generally larger in value and nearly constant. Note that the advantage of using $\mu_{\alpha_{\text {M}}}$ over the optimization scheme of the OT-RSH functional lies in its ease of evaluation.\footnote{The difference between the RS-DDH and the OT-RSH resides in the evaluation of the screening parameter; the former determined from linear response theory and the latter from an ionization potential tuning procedure.}

Table~\ref{table:gasphase-rsh} lists the evaluated vertical ionization potentials using the RS-DDH functional with $\mu=\mu_{\text{erfc-fit}}$ and the molecular polarizability screening parameter ($\mu_{\alpha_{\text{M}}}$) for the gas phase molecules that compose the solid molecular crystals investigated in Section II B. The vertical ionization potentials using $\mu_{\alpha_{\text{M}}}$ are in excellent agreement with experiment yielding a 2.3$\%$ MARE, whereas the electron affinities shown in Fig. S9 of the SM are described well for bound excited states ( MARE of 9.9$\%$). Using the screening parameter obtained from the complementary error function fit of the diagonal components of the dielectric matrix ($\mu_{\text{erfc-fit}}$) in the RS-DDH functional form to evalutate the vIP does not yield the same level of accuracy (MARE of 11.1$\%$). Also shown in Table~\ref{table:gasphase-rsh} are the vertical ionization potentials evaluated using the OT-RSH functional defined in Ref.~\onlinecite{RefaelyAbramson:2012uw}, which are of the same level of accuracy as the RS-DDH with $\mu=\mu_{\alpha_{\text{M}}}$. The small difference seen in the MAEs for RS-DDH($\mu_{\alpha_{\text{M}}}$) and OT-RSH is not statistically significant for the small set of molecules investigated (p-value of 0.18), indicating the two methods yield the same result.  

\begin{table}[]
\caption{The gas phase vertical ionization potential (vIP$_g$), in units of eV, of several molecules that compose the molecular crystals evaluated using the RS-DDH with $\mu=\mu_{\text{erfc-fit}}$ and $\mu=\mu_{\alpha_{\text {M}}}$, are shown in column 2 and column 3, respectively. Also shown are values determined using the OT-RSH functional defined in Ref.~\onlinecite{RefaelyAbramson:2012uw}, column 4. The experimental values (Exp.), taken from the NIST WebBook,\cite{Linstrom:2015} are listed in column 5.}
\label{table:gasphase-rsh}
\begin{tabular}{lcccc}
\hline
\hline
 &  \multicolumn{4}{c}{vIP$_g$ (eV)} \\
 &             RSH  & RSH & RSH & Exp. \\ 
 &              $\mu_{\text{erfc-fit}}$ & $\mu_{\alpha_{\text {M}}}$ & $\mu_{\text {OT-RSH}}$   \\ 
\hline
  C60         &  8.74 & 7.40&7.76 &7.60  \\
  PTCDA       &  9.23 & 8.20&8.20 &8.20  \\
  pentacene   &  7.08 & 6.29&6.29 &6.61  \\
  quinacridone&  8.44 & 7.35&7.35 &7.23\\
  tetracene   &  7.51 & 6.72&6.72 &6.97\\
  anthracene  &  8.14 & 7.30&7.30 &7.44 \\
  rubrene     &  7.28 & 6.28&6.16 &6.52\\
  napthalene  &  8.87 & 8.07&8.07 &8.14 \\
  benzene     &  10.15& 9.37&9.24 &9.25 \\
  ammonia     &  11.85&11.07&10.7 &10.8 \\
  acetic acid &  12.61&11.08&10.78&10.9 \\
  H$_2$O      &  13.78&13.01&12.55&12.62 \\
\hline
 ME (eV)      & 0.95 &-0.01 &-0.10& -- \\ 
 MAE (eV)     & 0.95 & 0.19 & 0.14& -- \\ 
 MRE ($\%$)   & 11.1 &-0.6  &-1.3 & -- \\ 
 MARE ($\%$)  & 11.1 & 2.3  & 1.9 & -- \\ 
\hline
\hline
\end{tabular}
\end{table}


\begin{figure}[]
\centering
\scalebox{0.42}{\includegraphics[trim= 100mm 10mm 80mm 20mm,clip, angle=0]{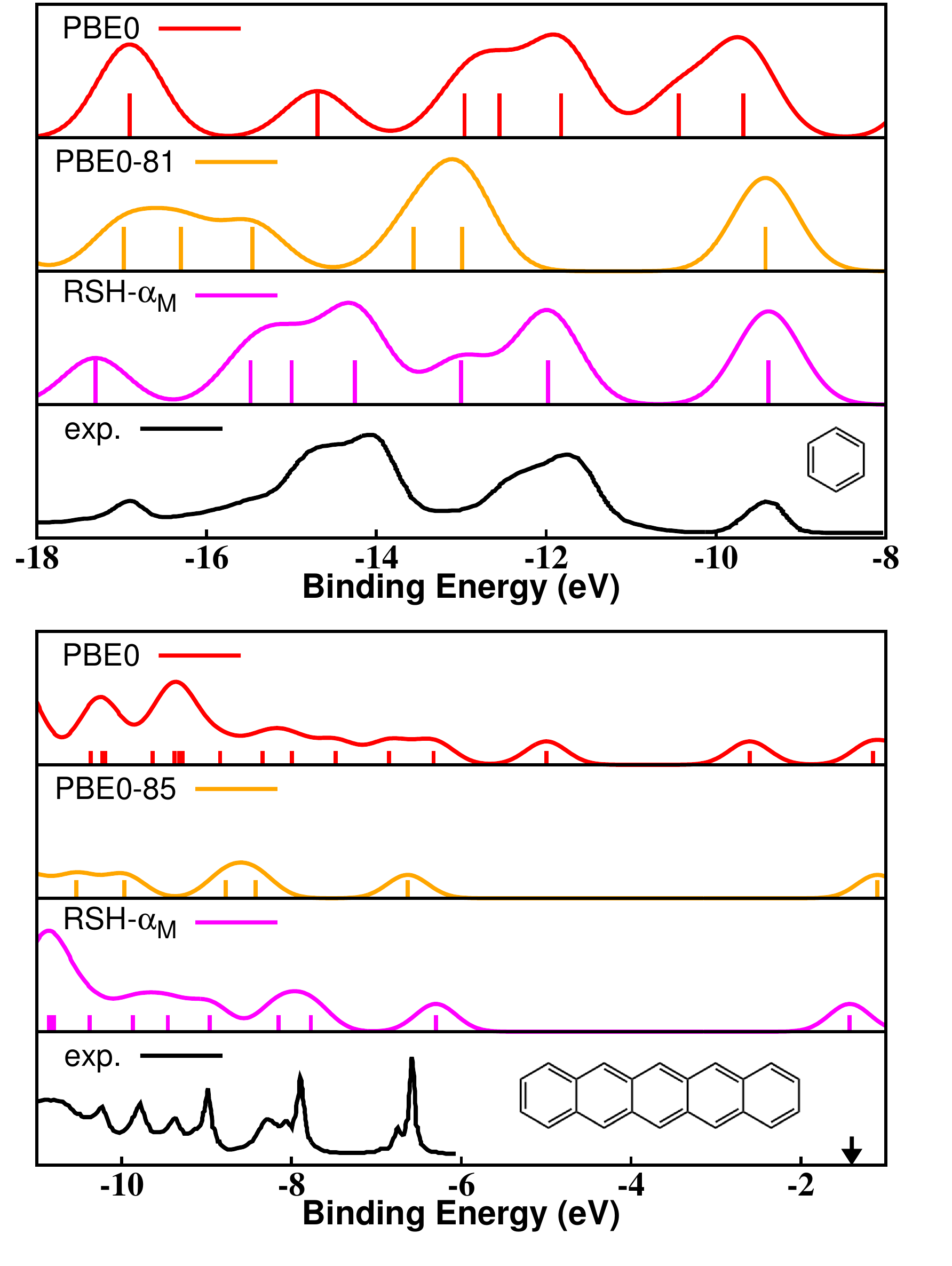}}
\caption{(Color online) The photoelectron spectrum of isolated benzene (top panel) and pentacene (bottom panel) molecules at several levels of theory; PBE0 (red), PBE$\alpha^*$\cite{Atalla:2013kk} labeled as PBE0-81, and PBE0-85, in the top and bottom panels respectively (orange), RSH-$\alpha_M$ (magenta), and experiment (black). The computed spectra are broadened by a Gaussian of width 0.38 eV and 0.28 eV for benzene and pentacene, respectively. The black arrow indicates the experimental electron affinity for pentacene. }
\label{fig:pes-molecules}
\end{figure}

We also computed the photoelectron spectra of several isolated molecules including benzene and pentacene, which are reported for various levels of theory in Fig.~\ref{fig:pes-molecules}, and compared to experiments. In the present work our focus is placed on the accurate determination of the spectra peak positions on an absolute scale rather than recovering the experimental peak intensities, and thus we did not formally compute intensities. The computed spectra shown are the density of states obtained by summing normalized gaussians centered at each KS energy state. For both molecules, the PBE0 spectra yield the poorest agreement with experiment. The highest occupied molecular orbital is best reproduced by the global hybrids PBE0-81 and PBE0-85, that minimize the difference between the KS eigenvalues and the GW quasi-particle corrections\cite{Atalla:2013kk}: 
\begin{equation}
\alpha^* = \text{arg min}_{\substack { \alpha}}|\langle \psi_{\text{H}} (\alpha)|\Sigma(\alpha) - v_{xc}(\alpha)|\psi_{\text{H}}(\alpha)\rangle|
\label{eqn:alpha-scheffler}
\end{equation}

However these functionals yield a poor description of the rest of the spectra. A similar observation was previously pointed out by K\"{o}rzd\"{o}rfer \textit{et al.}\cite{Korzdorfer:2012jx} in a study of tuned full-range hybrid functionals applied to molecules. Unlike full-range hybrid functionals, RSH introduces an effective spatially dependent screening that treats differently each KS state according to their spatial extent. This offers additional flexibility and thus yields further improvement in the spectral features over the full-range hybrid functionals. 

\section{ \textbf {Summary and Conclusions}}
\label{summary}
In summary, we defined a range separated (RS) form of  dielectric-dependent hybrid (DDH) functionals using material dependent, non empirical parameters and we showed that its performance is superior to that of non range-separated DDH functionals for the description of the electronic gap of a set of diverse inorganic semiconductors and insulators.  The same functional form also yields results in excellent agreement with experiments for the electronic gaps and vertical ionization potentials of molecular crystals, with a performance similar to that of full-range DDHs. 
Finally we presented a generalization of RS-DDHs to molecules and we discussed which parameters are appropriate to bridge long and short range components of the  generalized Kohn-Sham potential in the case of finite systems. The best agreement with experiments was obtained using a parameter related to the inverse of the cube root of the molecular polarizability, while the parameters used for extended systems (defined, e.g. using a Thomas-Fermi model fit to the diagonal components of the dielectric matrix) yielded less accurate results. 
We note in closing that the results presented here do not include the contribution from phonons coupling to the electronic states, which may be significant for light element solids.\cite{Cardona:2005ho,Giustino:2010ek,Antonius:2014im,Monserrat-Needs:2016} Further analysis of the gap renormalization due to the electron-phonon coupling will be reported elsewhere. 
\section*{ \textbf {Acknowledgments}}
We thank Bartolomeo Civalleri and the late Roberto Orlando for useful discussions regarding the technical details of the methodological implementations in CRYSTAL14. We also would like to acknowledge both Ikutaro Hamada and Ding Pan for their insightful discussions. This work was supported by NSF under the NSF center NSF-CHE-1305124 (J.H.S.) and by the Department of Energy grant DE-FG02-06ER46262 (M.G. and G.G.). 
An award of computer time was provided by the ASCR Leadership Computing Challenge (ALCC) program. This research used resources of the Argonne Leadership Computing Facility, which is a DOE Office of Science User Facility supported under Contract DE-AC02-06CH11357, resources of the National Energy Research Scientific Computing Center (NERSC), a DOE Office of Science User Facility supported by the Office of Science of the U.S. Department of Energy under Contract No. DE-AC02-05CH11231, resources of the Navy and Air Force DoD Supercomputing Resource Centers of the Department of Defense High Performance Computing Modernization Program, and resources of the University of Chicago Research Computing Center.

\bibliography{references-RSH}

\end{document}